\documentclass[aps,prl,reprint,groupedaddress]{revtex4-1}
\usepackage[colorlinks,linkcolor=blue, urlcolor=blue, anchorcolor=blue, citecolor=blue]{hyperref}

\usepackage{graphicx}% Include figure files
\usepackage{dcolumn}% Align table columns on decimal point
\usepackage{bm}% bold math
\usepackage{amsmath}
\usepackage{amsfonts}
\usepackage{color}
\pdfoutput=1
\draft % marks overfull lines with a black rule on the right

\begin{document}
	
	%\preprint{}
	
	\title{Water affects morphogenesis of growing aquatic plant leaves} %Title of paper
	
	
	
	\author{Fan Xu}
	%\thanks{Corresponding author}
	\email[Corresponding author.\\]{fanxu@fudan.edu.cn}
	%\homepage[]{http://homepage.fudan.edu.cn/fanxu/}
	\affiliation{Institute of Mechanics and Computational Engineering, Department of Aeronautics and Astronautics, Fudan University, Shanghai 200433, P.R. China}

    \author{Chenbo Fu}
	\affiliation{Institute of Mechanics and Computational Engineering, Department of Aeronautics and Astronautics, Fudan University, Shanghai 200433, P.R. China}
	%\thanks{Corresponding author}

    \author{Yifan Yang}
	\affiliation{Institute of Mechanics and Computational Engineering, Department of Aeronautics and Astronautics, Fudan University, Shanghai 200433, P.R. China}
	%\thanks{Corresponding author}
	
	
	\date{\today}
	
	\begin{abstract}
	Lotus leaves floating on water usually experience short-wavelength edge wrinkling that decays toward the center, while the leaves growing above water normally morph into a global bending cone shape with long rippled waves near the edge. Observations suggest that the underlying water (liquid substrate) significantly affects the morphogenesis of leaves. To understand the biophysical mechanism under such phenomena, we develop mathematical models that can effectively account for inhomogeneous differential growth of floating and free-standing leaves, to quantitatively predict formation and evolution of their morphology. We find, both theoretically and experimentally, that the short-wavelength buckled configuration is energetically favorable for growing membranes lying on liquid, while the global buckling shape is more preferable for suspended ones. Other influencing factors such as stem/vein, heterogeneity and dimension are also investigated. Our results provide a fundamental insight into a variety of plant morphogenesis affected by water foundation and suggest that such surface instabilities can be harnessed for morphology control of biomimetic deployable structures using substrate or edge actuation.
	\end{abstract}
	
	
	\maketitle%\maketitle must follow title, authors, abstract and \pacs
	
\begin{figure}[!htbp]
\includegraphics[width=8.8cm]{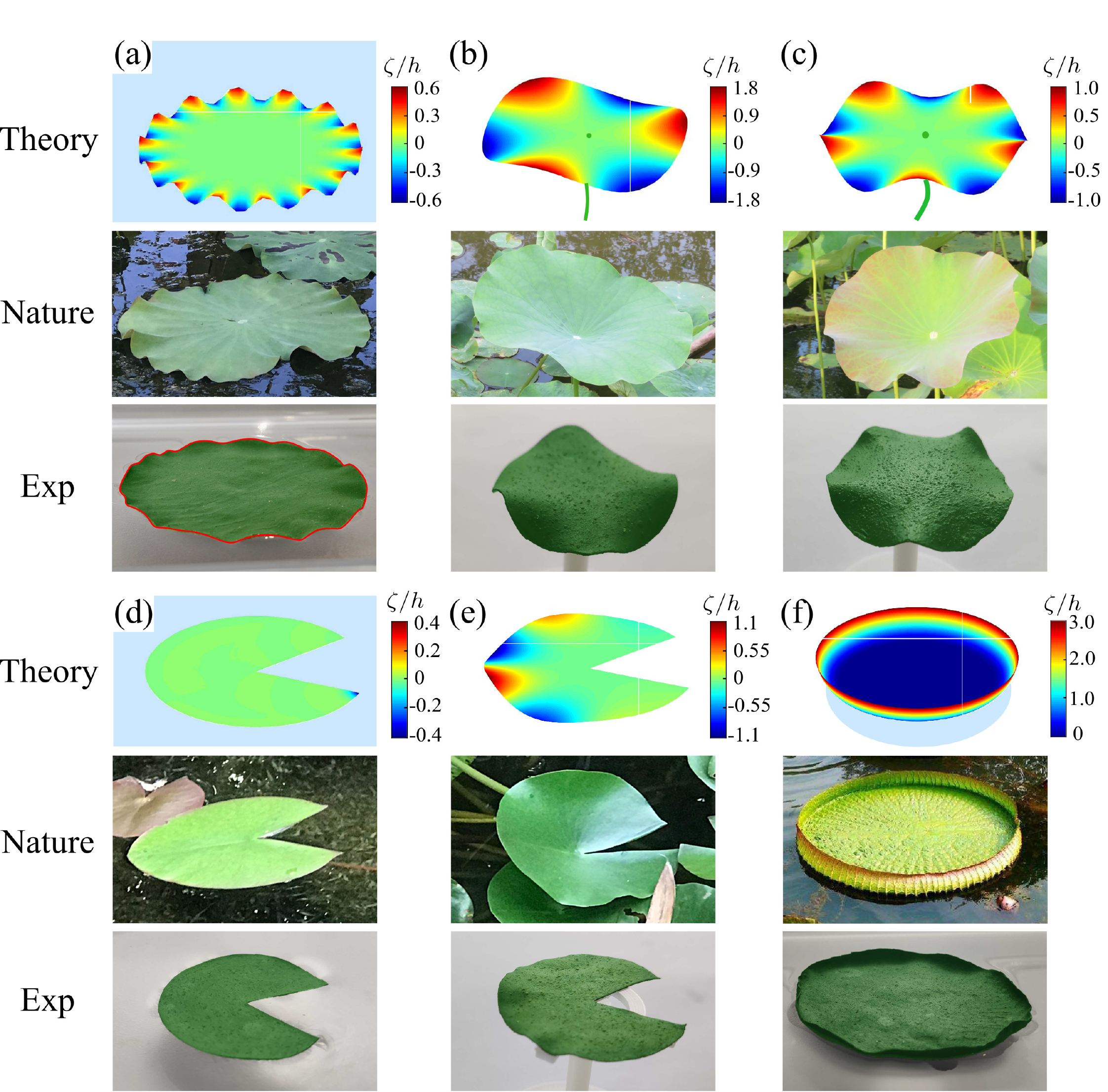}% Here is how to import EPS art
\caption{Water effect on morphogenesis of diverse aquatic plant leaves. A floating lotus leaf (a) grows with short waves along the edge (the wavy edge is highlighted by red color in experiments), while suspended lotus leaves (b) and (c) morph into long-wavelength ripples (depending on the size of main stem \cite{Sup}) near the margin. The more apparent global bending cone shape in natural lotus leaves might be caused by vein effect. With water foundation, the leaf of white water lily (d) remains flat, while for the suspended leaf (e) globally bends with long ripples near the edge. The leaf of Victoria water lily (f) where water substrate occupies about 80\% of the entire leaf area (local effect of liquid foundation) grows into a bowl-like shape with sharp edge bending. The blue background in (a), (d), and (f) represents water foundation. In theoretical calculations of (a)-(f), we took $R/h=50$ and $K_s h/E=1/1200$. The growth strain exponentially attenuates from the edge to the center, satisfying $\varepsilon^g=\epsilon^0\exp\left[-\tau (R-r)/{R}\right]$, in which $\tau=1$ and $\epsilon^0\sim 10^{-2}$.}
\label{circular}
\end{figure}

Having waves in morphological pattern is energetically favorable for thin living tissues such as leaves, flowers and biological membranes \cite{Sharon2004}, where spontaneous symmetry breaking induced by differential growth is normally considered as a significant factor in the origin of such complex patterns \cite{Audoly2003,Dervaux2008, Dervaux2009, Liang2009, Liang2011, Huang2018}, despite the known contribution of genes \cite{Sharon2002, Hill2005}. Growth-induced morphogenesis can be affected by many elements including intrinsic (\textit{e.g.}, gene \cite{Hill2005}) and external (\textit{e.g.}, phototropism \cite{Darwin1880}) ones. In this work, we observe that water can dramatically alter the morphogenesis of lotus leaves in the same plant, where the ones floating on water demonstrate flat geometry with short-wavelength wrinkles on the edge, while the leaves growing above water usually morph into a bending cone shape with long rippled waves near the edge, as shown in Fig. \ref{circular}. Such phenomenon reveals the interplay between internal growth-induced residual stresses and external support from the water (liquid substrate), which affects the morphogenesis of growing tissues. In this work, we explore, theoretically and experimentally, the water effects and other relevant factors that govern the differential growth induced pattern selection in diverse aquatic plant leaves.

We first establish a mathematical framework to study the underlying mechanism and to effectively predict the morphogenesis of growing lotus leaves. Growth is a complex process involving biochemical and physical reactions across different length and time scales. Since the growth process itself can be assumed to be very slow (takes time to relax to its equilibrium shape), the total deformation of the body can be viewed as time independent and is mainly attributed to the change of mass and elastic deformations. Before the deformation (reference configuration), the place of each material point is denoted by $\mathbf X$, while $\mathbf x$ represents the material point at the current configuration \cite{Sup}. We perform multiplicative decomposition of deformation gradient tensor $\mathbf{F}=\partial \mathbf x/\partial \mathbf X$ into elastic and growth parts \cite{Rodriguez1994}, namely $\mathbf F=\mathbf A \cdot\mathbf G$, where $\mathbf A$ is an elastic tensor characterizing the reorganization of the body, satisfying compatibility (no overlap) and integrity (no cavitation), while $\mathbf{G}=\mathbf{I}+\mathbf{g}$ (with $\|\mathbf{g}\|\ll 1$) is a growth tensor describing mass change, in which $\mathbf g$ denotes the gradient of displacement field $\mathbf u^*$ in the virtual configuration with zero deflection $\zeta=0$ \cite{Sup}. The Green-Lagrange strain tensor is then defined as $2\mathbf{E}=\mathbf{A}^T\cdot\mathbf{A}-\mathbf{I}$, in which $\mathbf{I}$ is the identity tensor. Considering the characteristic parameter $\varrho=h/L$ (thickness/length, with $\varrho \ll 1$) for thin leaves, dimensional analysis \cite{Sup} yields the F\"{o}ppl-von K\'{a}rm\'{a}n type strain coupled with spontaneous growth tensor,
\begin{equation}\label{a6}
\begin{split}
\displaystyle
\varepsilon_{ij}=&\displaystyle\frac 1 2\left(u_{i,j}+u_{j,i}+\zeta_{,i}\zeta_{,j}-g_{ij}-g_{ji}\right)\\
\displaystyle &-\zeta_{,\alpha}g_{\alpha,3}\delta_{i3}\delta_{j3}+\mathcal{O}\left(\varrho^3\right),
\end{split}
\end{equation}
where $u_{i, j}$ denotes the displacements, while $\zeta$ is the out-of-plane deflection of middle surface of the leaf. Latin indices $i, j,$... run from 1 to 3, while Greek indices $\alpha, \beta$,... take values in $\{1, 2\}$. A comma in subscript denotes a partial derivative and we use Einstein's convention for implicit summation on repeated indices. Since biological soft tissues hold a high volume fraction of water, they are elastically incompressible with $\det \mathbf{A}=1$.

We consider a lotus leaf floating on water as a thin homogeneous film lying on a substrate, and assume isotropy of Mooney-Rivlin constitution that follows a generalized Hooke's law for soft tissues \cite{Ogden1997,Wex2015}. The potential energy $\mathcal{P}$ of the system can be written as the sum of leaf part $\mathcal{P}_f$ and substrate part $\mathcal{P}_s$, namely $\mathcal{P}=\mathcal{P}_f+\mathcal{P}_s$. Inspired by finite elasticity theory of biological growth in soft tissues \cite{Dervaux2008, Dervaux2009}, we derive generalized F\"{o}ppl-von K\'{a}rm\'{a}n equations that can describe inhomogeneous differential growth of floating ($\mathcal{P}$) and free-standing leaves ($\mathcal{P}_f$) \cite{Sup}. Using the divergence theorem, the first variation of potential energy with respect to $\zeta$ and the in-plane elastic strains $\mathbf \varepsilon^0$ leads to the following equilibrium equations \cite{Sup}:
\begin{equation}\label{a20}
\begin{array}{l}
\displaystyle D\left( \Delta ^2\zeta -\Delta C_M \right) +\nabla \cdot \nabla \cdot \mathbf{M}^g-\nabla \cdot \left( h\mathbf{\sigma }^e\cdot \nabla \zeta \right) +K_s\zeta\\
=-\displaystyle\frac D2\nabla\cdot\nabla\cdot\left[ \begin{matrix}
g^0_{23,2}&-\displaystyle\frac{g^0_{13,2}+g^0_{23,1}}{2}  \\
-\displaystyle\frac{g^0_{13,2}+g^0_{23,1}}{2} &g^0_{13,1} \\
\end{matrix}\right], \\
\displaystyle \mathbf{\sigma}^e\cdot\mathbf{n}=\mathbf{0},
\end{array}
\end{equation}
where $D=Eh^3/9$ denotes the bending stiffness of leaves in which $E$ is Young's modulus, and $g^0_{\alpha 3}$ represents the growth strains of the neutral surface. The elastic stress $\mathbf{\sigma}^e$ is expressed as
\begin{equation}\label{a22}
\displaystyle
{\sigma}^e_{\alpha\beta}=\frac{2E}3\left[\left(\varepsilon_{\alpha\beta}^{0}-\varepsilon_{\alpha\beta}^{g}\right)+\left(\varepsilon_{\gamma \gamma}^{0}-\varepsilon_{\gamma \gamma}^{g}\right)\delta_{\alpha\beta} \right].
\end{equation}
For compatible growth, the terms $C_M$ and ${M}_{\alpha\beta}^g$ represent, respectively, the curvatures induced by the different horizontal and vertical growing rates of distinct leaf layers, which are respectively defined by
\begin{equation}\label{a23}
\begin{array}{l}
\displaystyle C_M=g_{\alpha 3,\alpha}^0,\\
\displaystyle {M}_{\alpha\beta}^g=\displaystyle\frac{2E}3\int^{h/{2}}_{-h/{2}}\left(\varepsilon_{\alpha\beta}^{g}+\varepsilon_{\gamma \gamma}^{g}\delta_{\alpha\beta} \right)X_3 \mathrm{d}X_3.
\end{array}
\end{equation}
Here, we assume a liquid substrate whose effective stiffness is given by $K_s=\rho \tilde{g}$, where $\rho$ is the liquid mass density and $\tilde{g}$ the gravitational acceleration \cite{Brau2013}. Therefore, the normal force acting on the leaf reads $\Sigma_{33}=-K_s\zeta$.
For a leaf growing freely on a surface, the governing equation (\ref{a20}) should satisfy the following boundary conditions:
\begin{equation}\label{a24}
\begin{array}{l}
\displaystyle \Delta\zeta-C_M-\displaystyle\frac{1}{2}\left(\zeta_{,\alpha\beta}-g^0_{\alpha 3, \beta}\right)l_\alpha l_\beta+\frac{M^g_{\alpha\beta}}{D}n_\alpha n_\beta=0,\\
\displaystyle \mathbf{n}\cdot\nabla\left[\left(\Delta\zeta-C_M\right)+\frac{M^g_{\alpha\beta}}Dn_\alpha n_\beta\right]\\+\displaystyle\frac 12\mathbf{l}\cdot\nabla\left[\left(\zeta_{,\alpha\beta}-\frac{g^0_{\alpha 3,\beta}+g^0_{\beta 3,\alpha}}{2}+\frac{4 M^g_{\alpha\beta}}D\right)l_\alpha n_\beta\right]=0,\\
\displaystyle \mathbf{\sigma}^e\cdot\mathbf{n}=\mathbf{0},
\end{array}
\end{equation}
in which $\mathbf{l}$ and $\mathbf{n}$ are tangential and normal unit vectors, respectively.

To solve growth-induced large deformations of the leaf/substrate model (\ref{a20}) and to trace the nonlinear morphological evolution of wrinkles, we adopt an efficient and robust numerical algorithm \cite{Sup} through coupling spectral collocation method \cite{Trefethen2000} for spatial discretization and \textit{Asymptotic Numerical Method} (ANM) \cite{Cochelin2007} for nonlinear resolution. Motivation of using spectral collocation approach stems from its global approximation nature, which can provide superior accuracy compared with finite difference and finite element method that are based on local arguments. The ANM is a numerical perturbation technique based on a succession of high-order Taylor series expansions with respect to an adaptive path parameter, which appears as a remarkably efficient continuation predictor to trace the post-buckling evolution on the equilibrium path. This resolution framework that combines advantages of both methods is particularly capable of solving highly nonlinear instabilities with finite deflection and deformation.

\begin{figure}[!htbp]
	\centering
	\includegraphics[width=9cm]{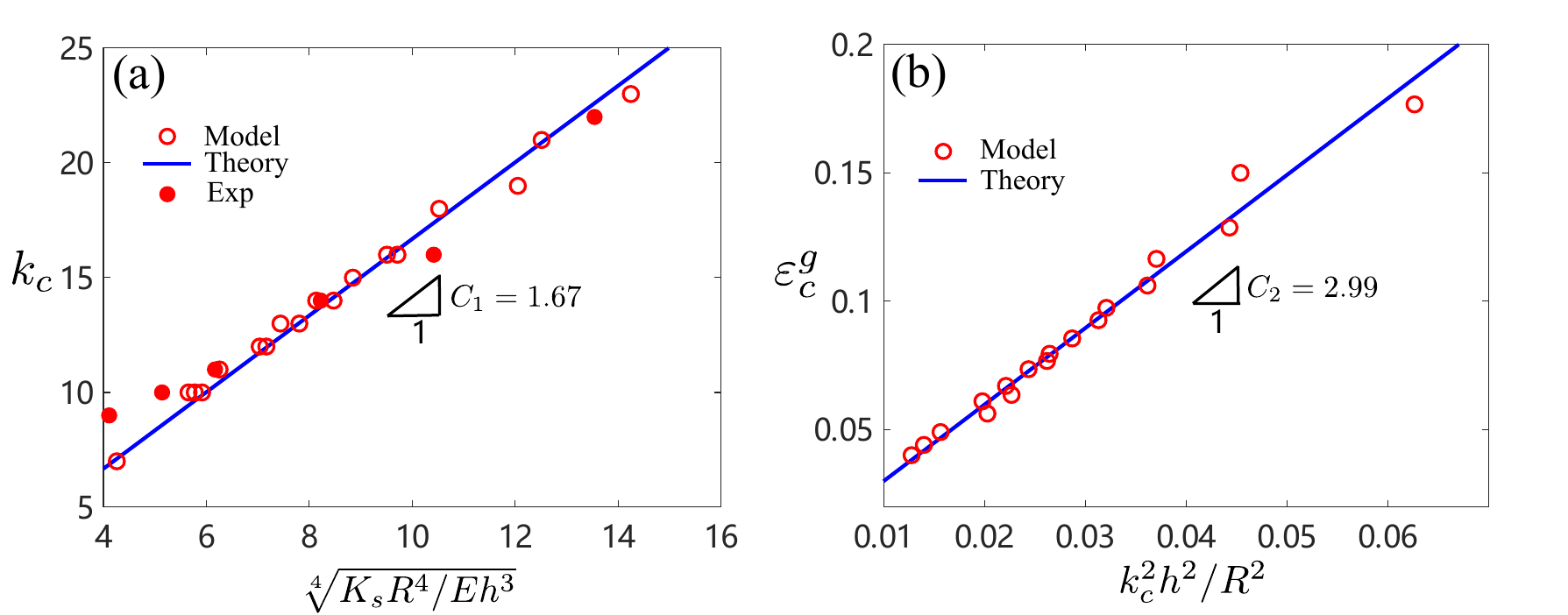}
	\caption{The comparison of model, theory, and experiments for water-foundation leaves: (a) Critical wrinkling wave number $k_c$ of circular floating lotus leaves as a linear function of $\sqrt[4]{K_s R^4 / E h^3}$. Our model and theoretical predictions agree well with experiments. (b) Critical growth strain $\varepsilon^g_c$ of circular floating lotus leaves as a linear function of $k_c^2 h^2/R^2$. Fitting coefficients: $C_1$ and $C_2$ \cite{Sup}.}
	\label{circu_line}
\end{figure}

\begin{figure}[!htbp]
\centering
\includegraphics[width=9cm]{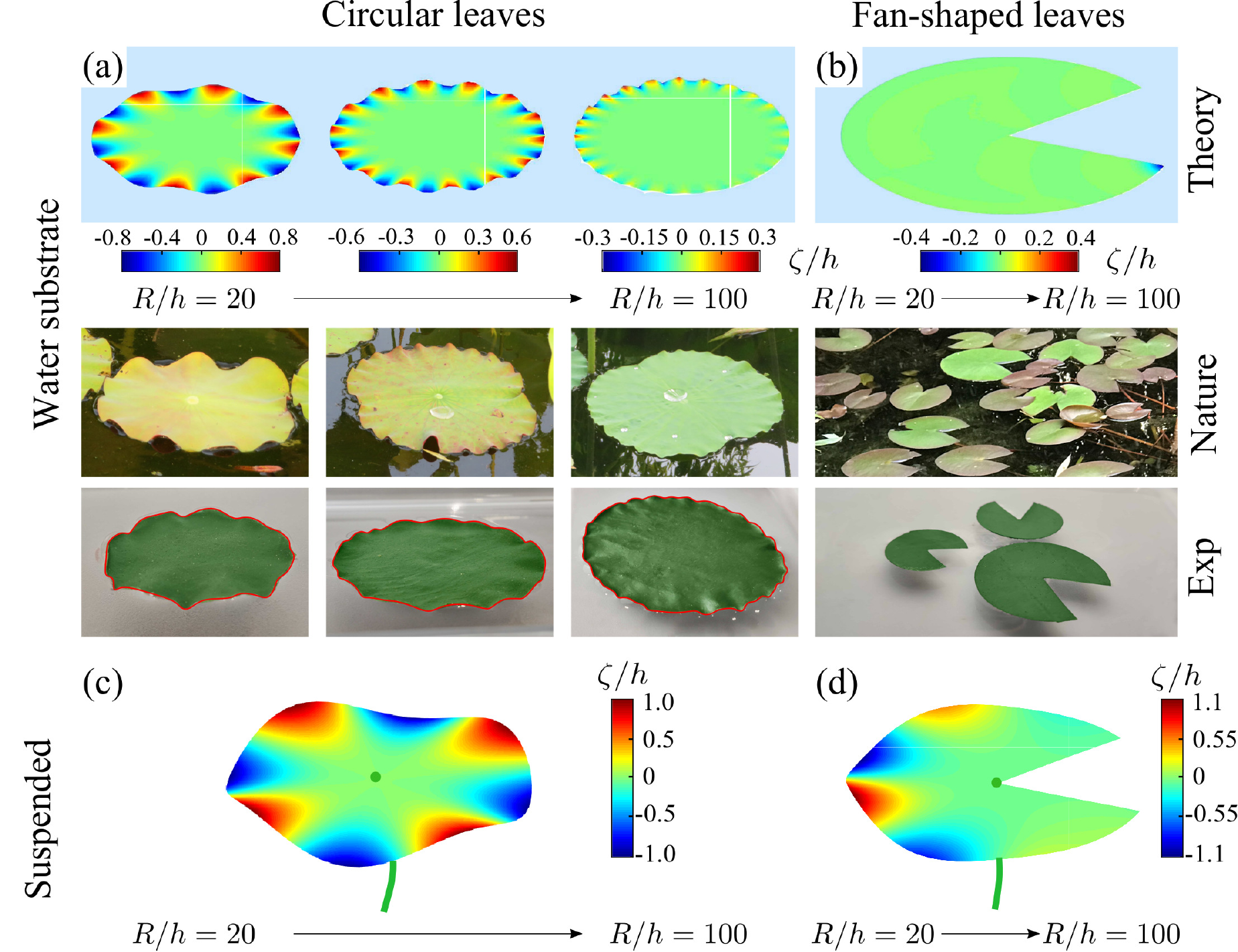}
\caption{Size effect ($R/h$) on growing shapes of lotus (circular) and white water lily leaves (fan-shaped): (a)-(b) water foundation (blue background), (c)-(d) suspended (stem support). For circular leaves floating on water, larger dimensionless radius $R/h$ leads to more waves, \textit{i.e.}, larger wave number (the wavy edges are highlighted by red color in experiments), while for circular leaves constrained by the central main stem, the dimension has no apparent effect on the wrinkling wave number and leaf morphology. Apart from size effect, this difference implies the important role played by water foundation in the morphology of growing aquatic plants. As for fan-shaped leaves, size effect, however, remains insignificant with/without water substrate.  Material parameter: $K_s h/E=1/1200$. Attenuation coefficient: $\tau=1$. Growth strain: $\epsilon^0\sim10^{-2}$.}
\label{size}
\end{figure}

Leaves and leaf-like structures exhibit a variety of patterns, ranging from the gracefully undulating submarine algal blades \cite{Koehl2008} to the saddle-shaped, coiled or edge-rippled leaves of many terrestrial and aquatic plants \cite{Sharon2007}. There exists a long-term misunderstanding in the literature, that is, only when the growth strain reaches a large value (\textit{e.g.}, $\sim$100\% in \cite{Liang2009}), the growing leaf can morph into edge ripples. This spontaneous strain, however, goes far beyond their validity range of linear elastic constitution. Here, we find that both saddle-shaped leaves and pleated leaves are related to their own structure and generalized constraints that they are subjected to. The origins of such constraints can be diverse such as water foundation and main stem of lotus leaves. We distinguish in Fig. \ref{circular} the different growth morphogenesis of lotus leaves which float on the water or suspend above the water. For a lotus leaf lying on the water, liquid substrate can dramatically affect its growth configuration with short-wavelength ripples along its edge yet remaining almost flat in the bulk (see Fig. \ref{circular}(a)), while a suspended lotus leaf above water tends to grow into long-wavelength ripples (see Fig. \ref{circular}(b)-(c)).

\begin{figure*}[!htbp]
	\includegraphics[width=17cm]{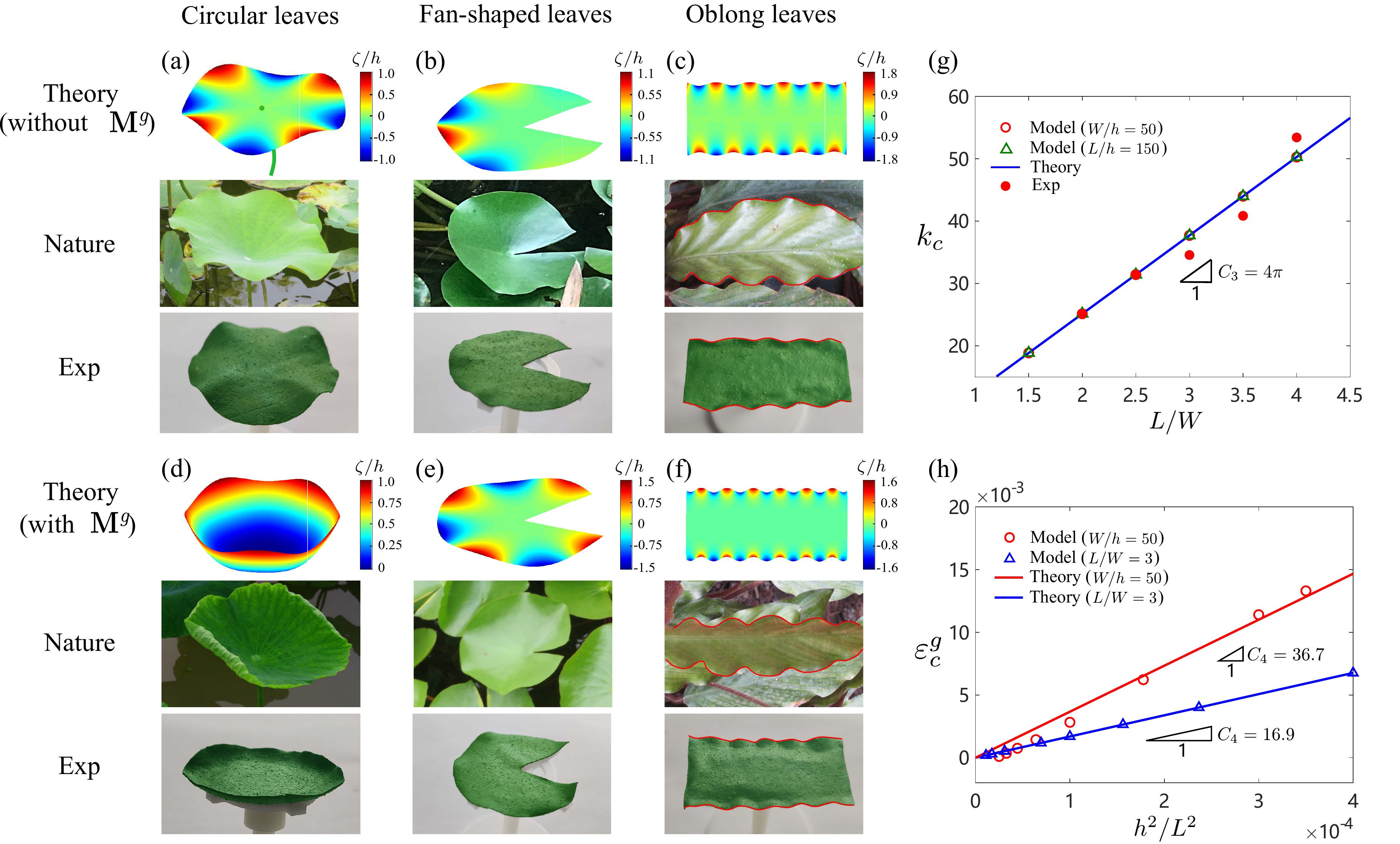}% Here is how to import EPS art
	\caption{Effect of growth-induced inhomogeneity on morphogenesis of diverse plant leaves. The upper line and lower line represent, respectively, our theoretical predictions of pattern selection without/by accounting for inhomogeneous growth induced bending moment. The columns classify the leaves by different geometries: circular, fan-shaped and rectangular. To distinguish between antisymmetric shape (c) and reflection-symmetric mode (f), the wrinkling edges are highlighted by red color in natural leaves and experiments. In theoretical calculations of (a)-(f), we took $R/h=50$ for circular and fan-shaped leaves, while $L/W=2$ and $W/h=50$ for oblong leaves. Attenuation coefficient $\tau=1$ and growth strain at leaf margin $\epsilon^0\sim 10^{-3}$ are set. (g) Critical wrinkling wave number $k_c$ of suspended oblong leaves as a linear function of aspect ratio $L/W$. Our model and theoretical predictions match well with experiments. (h) Critical growth strain $\varepsilon^g_c$ of suspended rectangular leaves as a linear function of $h^2/L^2$. Fitting coefficients: $C_3$ and $C_4$ \cite{Sup}.}
	\label{moment}
\end{figure*}

We further look into size effect on leaf morphogenesis. For circular lotus leaves floating on the water, larger dimensionless radius $R/h$ leads to larger wrinkling wave number due to dimension effect (see Figs.  \ref{circu_line}(a),  \ref{size}(a) and \cite{Sup}); whereas for suspended lotus leaves, the diameter of main stem/vein can select buckled wave number of shapes (see Fig. \ref{circular}(b)-(c)), instead of dimensionless size $R/h$ (see Fig. \ref{size}(c)). More precisely, the thicker the main stem is, the more waves are observed along the edge of leaves (see Fig. \ref{circular}(b)-(c)). Similar constraint influence can be observed in different geometry such as oblong leaves of \textit{Calathea rufibarba Fenzl} `Wavestar' (see Fig. \ref{moment} and \cite{Sup}). In addition, we find that the critical growth strain of the constrained leaf by the main vein remains smaller than that of the unconstrained leaf, which implies why edge-rippled morphogenesis other than saddle shape is energetically favorable for long leaves in nature \cite{Sup}. For oblong leaves with a fixed width/thickness ratio, the number of waves along the edge appears to be proportional to the aspect ratio (see Fig. \ref{moment}(g) and \cite{Sup}).

Water effect can also be found in leaves with different geometries such as a fan-shaped leaf of white water lily. In addition to its stiffer material property to lotus leaf, the geometric defect can relieve compressive stresses and thus this leaf experiences distinct growing morphology. With water support, the natural growth strain cannot reach its critical wrinkling value and thus the leaf remains flat (see Fig. \ref{circular}(d)), while for the suspended leaf with lower structural stiffness, it can globally deform with slight ripples near the edges (see Fig. \ref{circular}(e)). This phenomenon is found to be universal and size-independent (see Fig. \ref{size}(b) and (d)). An interesting phenomenon of water effect exists in leaves of Victoria water lily where water substrate occupies about 80\% of the entire leaf area (local effect of liquid foundation). Both observation and theory in Fig. \ref{circular}(f) show that the leaf of Victoria water lily morphs into a bowl-like configuration with a sharp bent edge upon growth.

To further explore growth-induced critical instability conditions of leaf morphogenesis, we derive a scaling law to predict the buckling wave number and growth strain \cite{Sup}. As an ansatz, we consider the following forms for the displacements upon wrinkling: $\zeta=A(kr/R)\cos(k\theta)$, where $k$ denotes the wave number, and the wrinkling amplitude, $A(kr/R)$, is a decaying function that can be approximated by an exponential law. Minimization of the system yields the critical buckling wave number and growth strain of circular floating lotus leaves that obey the following linearized relations, confirmed by both computations based on our models and experiments (see Fig. \ref{circu_line}):
\begin{equation}\label{a84}
  \displaystyle k_c \sim \sqrt[4]{\frac{K_sR^4}{Eh^3}},\quad \varepsilon^g_c \sim \frac{h^2k_c^2}{R^2}.
\end{equation}
More details on theoretical derivations can be found in \cite{Sup}.

Growing soft tissues usually exhibit spatial inhomogeneities and anisotropic growth upon large deformation \cite{Coen2004}, partially because of spatial heterogeneous distribution and transport of nutrient sources. A simple and realistic way to consider this heterogeneity is to take an exponential decay law for in-plane differential growth ($\varepsilon^g_{\theta\theta}/\varepsilon^g_{rr}$ and $\nabla\varepsilon^g_{\alpha\beta}$ on growth morphology). Here, we consider a growth strain attenuated from the edge to the center, satisfying $\varepsilon^g_{\alpha \beta}=\epsilon^0_{\alpha \beta}\exp[-\tau(R-r)/R]$, in which $\tau$ is an attenuation coefficient. The increase of heterogeneity of growth strain leads to more rippled edges, while it reduces into the homogeneous growth with flattened margin when $\tau=0$ \cite{Sup}. We find that with the increase of growth anisotropy $\epsilon^0_{\theta\theta}/\epsilon^0_{rr}$, the lotus leaf deforms into wavier shape (yet with a limit value). However, the spatial inhomogeneity of growth strain (in-plane attenuation) has no significant influence on leaf morphology \cite{Sup}. We further investigate the out-of-plane heterogeneity along the thickness direction caused by phototropism of the plants, \textit{i.e.}, leaves generally grow faster on the back side than the side facing the sun light \cite{Darwin1880}. This inhomogeneity can result in a growth-induced moment $\mathbf M^g$ that can greatly alter the morphogenesis of leaves. In Fig. \ref{moment}, we demonstrate the influence of out-of-plane growth heterogeneity on pattern formation of diverse leaves with different geometries. For circular leaves such as lotus, the growth moment leads to a distinct coupled behavior of global bending and edge wrinkling. However, for fan-shaped leaves such as white water lily, effect of growth curvature remains rather limited, probably due to geometric imperfection. For oblong leaves such as \textit{Calathea rufibarba Fenzl} `Wavestar', we find that the growth curvature determines the symmetry selection of wrinkling morphology. Without a heterogeneous growth-induced moment, an oblong leaf prefers to buckle into an antisymmetric shape, whereas with growth curvature, reflection-symmetric pattern is more favorable. Such pattern selection is attributed to the growth moment that can impose a symmetric bending disturbance, towards triggering a reflection-symmetric wavy shape. Dimensional analysis \cite{Sup} yields the critical buckling wave number $k_c\sim{L}/{W}$ and growth strain $\varepsilon^g_c\sim h^2/{L^2}$, consistent with our model predictions, as shown in Fig. \ref{moment}(g)-(h).

Inspired by the substrate-affected biological growth, we next design a demonstrative experiment to harness such mechanism for pattern formation, by using a water-swelling rubber (WSR) that can hold maximum volume swelling rate over 300\% after sufficient water absorption \cite{Sup}. We first compress the WSR into thin sheets with a minimum thickness $\sim0.3$ mm, and then cut them into different leaf-like shapes. Interestingly, these thin sheets floating on water morph into similar patterns as in nature and predicted by our theoretical models (see Figs. \ref{circular} and \ref{size}). For suspended leaves, we design 3D printed hollowed structures to support the samples \cite{Sup}. Such a simple setting can reduce the gravity effect on pattern formation upon water absorption (see Figs. \ref{circular} and \ref{moment}). We design a simple test to explore water-affected pattern transition \cite{Sup}. The buckling wavelength of a floating sheet taken out of water increases, while that of a suspended one put on water decreases. This fact suggests that suspended and floating shapes can interconvert into one another. Such interconversion, indeed, proves the interplay between internal growth-induced stresses and external support from the water (liquid substrate), which significantly affects the morphogenesis of growing tissues. In addition, we can flexibly program the water absorption area on the sheet surface to mimic various growth conditions, such as homogeneous/inhomogeneous in-plane growth, growth-induced moment and partial liquid foundation, in order to realize targeted morphological patterns \cite{Sup}. Our experiments not only reproduce the diverse morphogenesis of natural leaves (consistent with our theoretical predictions), but also shed light on designs of wrinkle-tunable multifunctional membrane surfaces and structures, based on our fundamental understandings and models.

In summary, we have revealed diverse growth-induced morphogenesis and pattern evolution of terrestrial and aquatic plant leaves such as lotus and water lilies, which can be well predicted by our theory, in good agreement with carefully designed experiments. A remarkable finding lies in the water (liquid substrate) effect on the shape selection of leaves. Precisely, leaves floating on the water exhibit short-wavelength edge wrinkling that decays towards the center, while the ones growing above the water usually morph into global bending cone shape with long rippled waves near the margin. Notably, leaves of Victoria water lily with partial water support (local effect) are prone to grow into a bowl-like shape. Besides, other influencing factors, such as mechanical constraints from the stem/vein, heterogeneity-induced growth curvature and size effect, can alter the shape of leaves. Understanding growth-triggered morphological evolution and in particular the dependence of wrinkling behavior on liquid foundation can help design biomimetic deployable structures that quantitatively harness surface instabilities using substrate or edge actuation.
	
	This work is supported by the National Natural Science Foundation of China (Grants No. 11602058, 11872150, 11772094 and 11890673), Shanghai Rising-Star Program (Grant No. 19QA1400500), Shanghai Chenguang Program (Grant No. 16CG01), and State Key Laboratory for Strength and Vibration of Mechanical Structures (Grant No. SV2018-KF-17).
	
	%\section{}
	%%\label{}
	%\subsection{}
	%\subsubsection{}
	
	% If in two-column mode, this environment will change to single-column format so that long equations can be displayed.
	%\begin{widetext}
	%$$\mbox{put long equation here}$$
	%\end{widetext}
	
	% Figures should be put into the text as floats.
	% Use the graphics or graphicx packages (distributed with LaTeX2e).
	% Put the label that you will use with \ref{} command in the braces of the \label{} command.
	%
	% \begin{figure}
	% \end{figure}
	
	% \begin{table}
	% \end{tabular}
	% \end{table}
	
	%\begin{acknowledgments}
	% Put your acknowledgments here.
	%\end{acknowledgments}
	
	\bibliography{prl}
	
\end{document}